\documentclass[aps,pre,twocolumn,10pt,
 notitlepage,
 floatfix,
 nofootinbib,
 superscriptaddress]{revtex4-2}

\usepackage{graphicx}
\usepackage{dcolumn}
\usepackage{bm}

\usepackage[colorlinks=true,
            linkcolor=blue,
            citecolor=blue,
            urlcolor=blue]{hyperref}

\usepackage[T1]{fontenc}
\usepackage[utf8]{inputenc}
\usepackage{microtype}
\usepackage{amsmath}
\usepackage{amssymb}
\usepackage{orcidlink}
\newcommand{\av}[1]{\left\langle {#1}\right\rangle}

\begin{document}

\title{Intermittent quarantine suppresses epidemic spreading beyond simple contact reduction}

\author{Juliane T. Moraes\orcidlink{0000-0002-9199-8237}}
\affiliation{Departamento de Física, Universidade Federal de Viçosa,
36570-900, Viçosa, MG, Brazil}
\affiliation{Instituto de Ciência e Tecnologia, Universidade Federal de São Paulo,
12247-014, São José dos Campos, SP, Brazil}

\author{Silvio C. Ferreira\orcidlink{0000-0001-7159-2769}}
\affiliation{Departamento de Física, Universidade Federal de Viçosa,
36570-900, Viçosa, MG, Brazil}
\affiliation{Instituto de Ciências Matemáticas e de Computação,
Universidade de São Paulo, São Carlos, SP 13566-590, Brazil}
\begin{abstract}
Social interactions are shaped by networks that can be highly heterogeneous, with some individuals potentially interacting with many others. However, physical and temporal constraints limit the number of simultaneous interactions, making social contacts inherently intermittent. Here, we investigate how intermittency affects epidemic spreading on complex networks considering two distinct mechanisms: intermittency of connections (links) and individuals themselves (nodes), with periods of activity and inactivity governed by renewal processes with broadly distributed inter-event times. Combining mechanistic theoretical approaches with stochastic simulations, we show that these two forms of intermittency have fundamentally different effects on epidemic spreading. Link intermittency largely preserves the mechanisms responsible for sustaining transmission, and epidemic behavior can be largely explained by the average reduction in contact opportunities. Contrastingly, node intermittency is substantially more effective at containing spreading because the simultaneous interruption of all contacts suppresses transmission while recovery continues independently. The non-Markovian nature of node intermittency plays a key role: shorter and less heterogeneous inter-event times increase epidemic thresholds and enhance epidemic containment. Our results highlight how the temporal organization of individual activity, beyond the average reduction in contacts, can substantially alter epidemic spreading and may help inform interventions that exploit intermittency in social and biological contact networks.
\end{abstract}

\maketitle

\section{Introduction}

The spread of infectious diseases is shaped by interactions among individuals, which can be represented by networks where nodes correspond to individuals that can carry and transmit a pathogen and links represent the contacts through which transmission may occur~\cite{Keeling1999,Lloyd2001,Pastor-Satorras2001}. Much of our theoretical understanding of epidemic spreading has been developed using static networks, implicitly assuming that the structure of contacts changes on timescales much longer than the spreading process itself~\cite{Pastor-Satorras2015}. However, contacts in real populations are inherently dynamic: individuals move, alternate between periods of activity and inactivity, and continuously establish and interrupt interactions. Since the timing and duration of contacts are crucial for transmission, temporal-network approaches have been extensively developed to account for the dynamical nature of interactions~\cite{Holme2012,Masuda2016}.

Temporal networks represent interactions as sequences of contact events, retaining information about when contacts occur and how long they last~\cite{Masuda2016}. Empirical studies of face-to-face interactions reveal a highly heterogeneous temporal organization: many contacts are short and occur close together in time, whereas others last much longer or are separated by extended periods without interaction~\cite{Cattuto2010,Stehle2011,Fournet2014}. Such temporal heterogeneity is closely related to the bursty nature observed in many forms of human activity, where periods of intense activity alternate with long periods of inactivity~\cite{Barabasi2005}. These broad distributions of waiting times can have important consequences for spreading processes, including anomalously slow propagation at long times~\cite{Min2011}. Thus, potential transmission opportunities are far from evenly distributed in time, and their temporal organization can substantially alter the course of a spreading process.

A substantial body of work has investigated these effects using models in which interactions are represented as instantaneous contact events, during which transmission occurs with a given probability, while recovery from the infectious state proceeds independently of these contacts~\cite{Perra2012,Holme2015a,Rocha2013}. Such an event-based representation is particularly natural when transmission opportunities are associated with discrete encounters, as in networks of sexual contacts~\cite{Holme2015a}. An important example is provided by activity-driven networks~\cite{Perra2012}, in which each individual is assigned an intrinsic activity rate $a_i$ that determines its propensity to initiate contacts. Active individuals establish temporary connections that are subsequently removed, continuously reshaping the instantaneous contact network. Heterogeneity in individual activity strongly affects the conditions for sustained transmission and, consequently, the epidemic threshold. More broadly, explicitly accounting for the temporal ordering of contacts can lead to epidemic outcomes substantially different from those predicted by static or time-aggregated representations of the same system~\cite{Holme2012,Holme2015a}.

An alternative approach, which we follow in the present work, considers an underlying backbone network that remains fixed while its elements switch between active and inactive states according to stochastic processes. In particular, temporal-network models have been developed in which each edge alternates between active and dormant periods, whose durations can follow broad, non-exponential distributions~\cite{Han2023a,Han2023b}. This representation separates the relatively stable structure of potential interactions---who can interact with whom---from their temporal availability for transmission. These studies have shown that the mean durations of active and dormant periods directly affect the epidemic threshold, whereas the detailed temporal statistics can influence the transient spreading dynamics~\cite{Han2023a,Han2023b}. More generally, whether temporal and non-Markovian effects can be reduced to effective Markovian spreading on a static network depends on the specific mechanism governing contact and transmission events~\cite{Han2023a}.

The works of Han and collaborators~\cite{Han2023a,Han2023b} follow an edge-based approach, in which the temporal component of the network is introduced by allowing individual edges to alternate between active and inactive states according to renewal processes. The durations of these states can be drawn from broad distributions, thereby incorporating the heterogeneous temporal patterns observed in human contacts. An alternative class of temporal-network models assigns the temporal dynamics to individuals rather than to individual connections~\cite{Perra2012,Rocha2013}. In activity-driven models, for example, individuals become active according to node-specific activity rates and establish temporary contacts while active~\cite{Perra2012}, whereas models based on bursty vertex activation describe individuals generating contacts according to heterogeneous inter-event-time distributions~\cite{Rocha2013}. In both cases, however, individual activity generates the instantaneous network of contacts rather than modulating the availability of connections on a pre-existing contact network.

The COVID-19 pandemic brought unprecedented attention to interventions aimed at limiting disease transmission and preventing health-care systems from becoming overwhelmed~\cite{Flaxman2020,Maier2020}. Among the most stringent measures, lockdowns and mobility restrictions substantially reduced opportunities for potentially infectious contacts, contributing to the containment of epidemic growth~\cite{Flaxman2020,Maier2020}. Such interventions, however, came at substantial social and economic costs and cannot be sustained indefinitely. This raises a broader question of both theoretical and practical relevance: given a reduction in overall contact activity, does it matter how this reduction is distributed across individuals and their interactions?

Here, we advance the understanding of non-Markovian temporal networks by systematically comparing the effects of edge and node intermittency on epidemic spreading over the same fixed underlying network. Activation and deactivation of either edges or nodes are governed by stochastic renewal processes, independently of the epidemic states of the individuals involved. This assumption is particularly suitable for epidemiological scenarios involving asymptomatic or presymptomatic transmission~\cite{Li2020}, in which infected individuals may continue their usual patterns of activity and inadvertently transmit the pathogen to susceptible contacts. To disentangle the effects of contact intermittency from those arising from the epidemic dynamics itself, we consider the paradigmatic susceptible-infected-susceptible (SIS) and susceptible-infected-recovered (SIR) processes with Markovian transmission and recovery. Thus, all non-Markovian effects investigated here originate exclusively from the temporal organization of contacts. We show that edge and node intermittency lead to fundamentally different outcomes. While the effects of edge intermittency can be effectively reduced to those of the corresponding Markovian dynamics on the static underlying network through a suitable rescaling of the transmission rate, node intermittency cannot be captured by such a simple reduction and suppresses epidemic spreading substantially more efficiently than expected from the average reduction in contact availability alone. We rationalize this difference in terms of the distinct microscopic mechanisms that drive epidemic activation on heterogeneous networks. Finally, we investigate how these results depend on broad inter-event-time distributions and on the heterogeneity of the underlying network structure. We find that shorter and less heterogeneous inter-event times lead to higher epidemic thresholds, demonstrating that not only the average availability of contacts but also their temporal organization plays a key role in epidemic containment.

\section{Model and methods}

\begin{figure*}[t]
	\centering
	\includegraphics[width=0.9\linewidth]{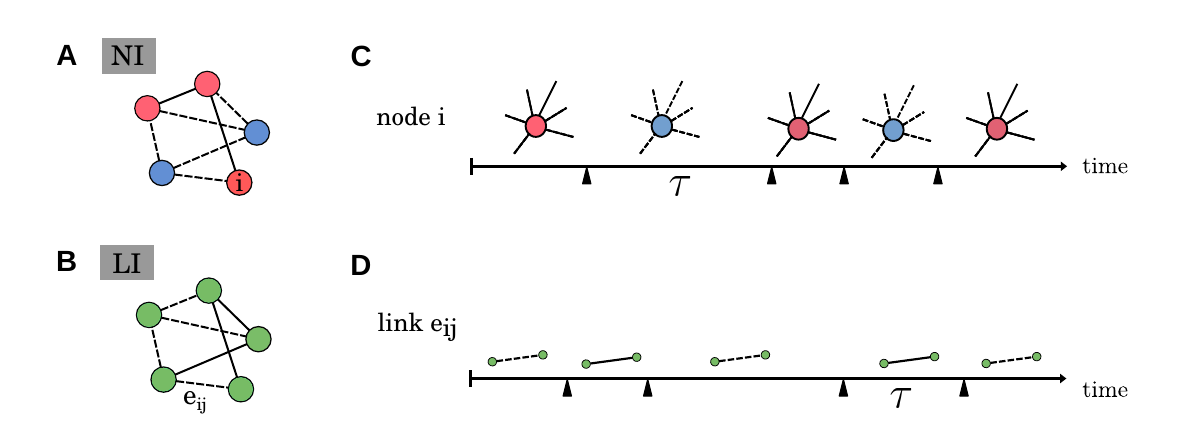}
	\caption{
       Intermittent network model. Independent renewal processes with distributions $\psi_\text{on}(\tau)$ and $\psi_\text{off}(\tau)$ governing activation and deactivation, respectively, are associated with each node or link of a static backbone network, leading to (A) node intermittency (NI) and (B) link intermittency (LI). Active and inactive nodes (links) are represented by red (solid) and blue (dashed) circles (lines), respectively. Sequences of activation and deactivation events are illustrated for (C) NI and (D) LI, with changes in the state of a node or link indicated by black arrows.
    }
	\label{fig:niandeischeme}
\end{figure*}

Individuals are represented by the nodes of a static complex network~\cite{Pastor-Satorras2015}: the backbone network, while edges represent potential interactions through which transmission can occur. The population size (number of nodes) is constant and denoted by $N$. While the backbone is static,  either nodes or links can switch between active (on) and inactive (off) states. When the elements are active, the epidemic dynamics evolves as if it was running on the static backbone network~\cite{Cota2017} while inactive elements temporarily suppress the associated transmission pathways. Two models are investigated. For \textit{link intermittency} (LI), contagion through a link is allowed only when it is active. Similarly,  for \textit{node intermittency} (NI), an infected node who are not active cannot transmit to nor be infected by any of its neighbors. We consider link and node intermittency separately.

Independent renewal processes\footnote{A renewal process describes a sequence of events in which, after each event, the waiting time until the next one is independently drawn from a prescribed probability distribution~\cite{Cox2017}.} are assigned to each link or node, depending on the intermittency model: an active element becomes inactive after a random time $\tau$ drawn from the probability density function (p.d.f.) $\Psi_{\textbf{on}}(\tau)$. Similarly, an inactive element is activated after a time drawn from $\Psi_{\textbf{off}}(\tau)$. Figure~\ref{fig:niandeischeme} illustrates node and link intermittencies. Motivated by the burst patterns observed in human activity and contact dynamics~\cite{Karsai2018,Cattuto2010,Barabasi2005,Fournet2014}, we consider broadly distributed waiting times with a power-law tail proportional to $\tau^{-\beta}$. Such distributions introduce memory into the temporal network, and the dynamics are non-Markovian\footnote{For exponentially distributed waiting times, the switching process is Markovian and memoryless: the probability of changing state does not depend on how long the element has already remained in its current state~\cite{Masuda2016}.}: the probability that an element changes state depends on how long it has already remained in its current state. As in Ref.~\cite{Han2023a}, we use Lomax-type distributions for both active and inactive periods,
\begin{equation}
	\Psi_\text{x}(\tau) = \frac{(\beta_\text{x}-1)(\tau_0 + a_\text{x})^{\beta_\text{x}-1}}{(\tau + a_\text{x})^{\beta_\text{x}}}, \quad \tau > \tau_0 \ ,
	\label{eq:Psi}
\end{equation}
where the subscript $\text{x}$ refers to ``on'' or ``off''. Here, $\beta_\text{x}$ controls the heterogeneity of the waiting times, while the parameter $a_\text{x}$ allows $\beta$ to be varied at fixed mean waiting time,
\begin{equation}
	\av{\tau}_\textbf{x} = \frac{a_\text{x} + \tau_0(\beta_\text{x} -1)}{\beta_\text{x} -2}, \quad \beta_\text{x} > 2.
\end{equation}

All simulations start when the temporal network is stationary. This can be achieve as follows~\cite{Han2023a}: Links or edges are labeled as on with probability
$
	p_\textbf{on} = {\av{\tau}_\text{on}}/{(\av{\tau}_\text{on}+\av{\tau}_\text{off})},
$
and off otherwise. The p.d.f of the time since the edge entered the on-state is~\cite{Han2023a}
\begin{equation}
	\Phi_\textbf{on}(\tau) = \frac{\int_\tau^\infty \psi_\text{on}(\tau') d\tau'}{\av{\tau}_\textbf{on}}.
\end{equation}

We investigate the two paradigmatic epidemic processes, SIS and SIR, using continuous-time dynamics~\cite{Pastor-Satorras2015}. Nodes can be susceptible, infected, or recovered. Susceptible nodes become infected at rate $\lambda$ per active contact with an infected node, while infected nodes recover spontaneously at rate $\mu$. Upon recovery, nodes become permanently immune in the SIR model or susceptible again in the SIS model. The epidemic dynamics are simulated using the optimized Gillespie algorithm (OGA)~\cite{Cota2017}, with inactive links or nodes excluded from transmission events. Since the network dynamics evolve independently of the epidemic states, the simulation of the intermittency processes can be optimized, as detailed in Appendix~\ref{app:algorithm}.

The SIS model exhibits an active steady state when the infection rate $\lambda$ exceeds the epidemic threshold $\lambda_c$. In a finite population, however, random fluctuations can eventually eliminate all infected individuals, even above the epidemic threshold. Once this disease-free state is reached, the epidemic cannot restart spontaneously, and the dynamics remains trapped there. To study the sustained epidemic regime without this finite-size limitation, we use a quasi-stationary (QS) method, in which the dynamics is reactivated whenever the disease-free state is reached.\footnote{We are interested in the large-population limit, for which the QS procedure does not affect the relevant epidemiological quantities~\cite{Sander2016}. Below the epidemic threshold, the number of infected nodes maintained by the reactivation mechanism remains finite, such that the prevalence $\rho\sim 1/N\rightarrow0$ as $N\rightarrow\infty$. Above the threshold, the probability of reaching the disease-free state vanishes in this limit.} Specifically, we use the hub-reactivation QS method~\cite{Sander2016}: whenever the disease-free state is reached, the most connected node of the backbone network is reinfected and activated, restarting its activity cycle. We have verified that our results are robust with respect to different QS methods.

\begin{figure*}[!t]
	\centering
	\includegraphics[width=0.42\linewidth]{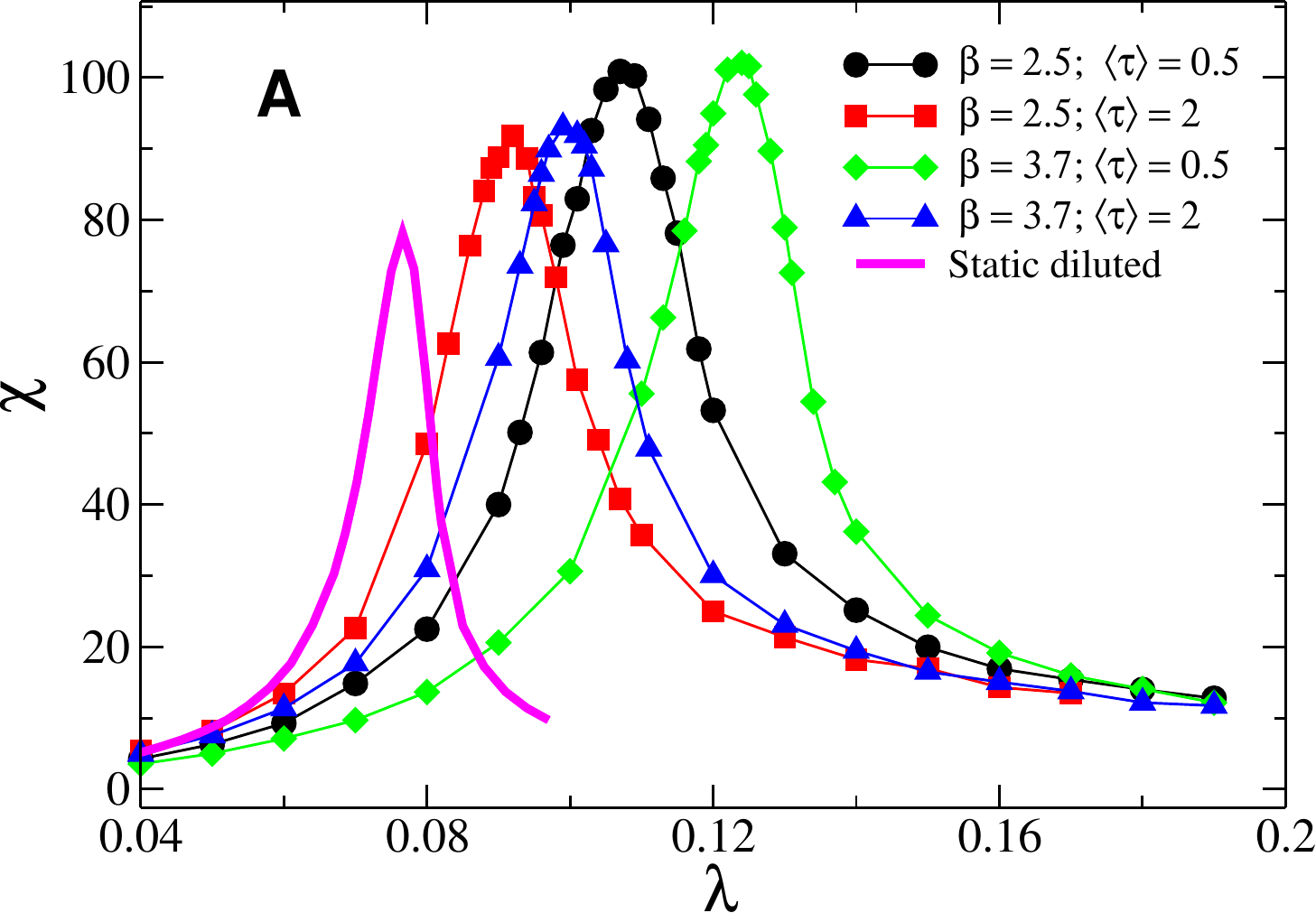} ~~ 
	\includegraphics[width=0.4\linewidth]{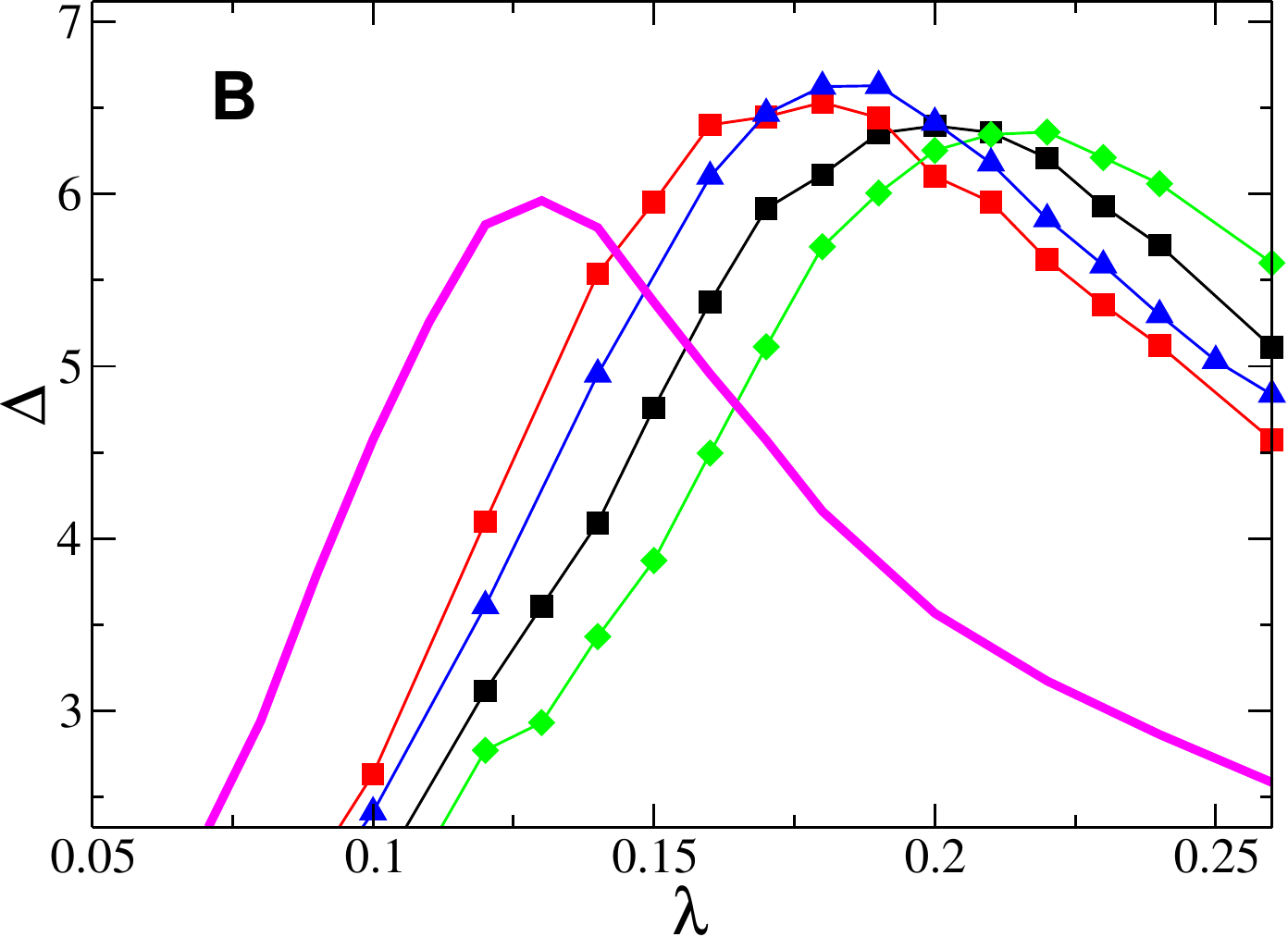}\\
	\includegraphics[width=0.42\linewidth]{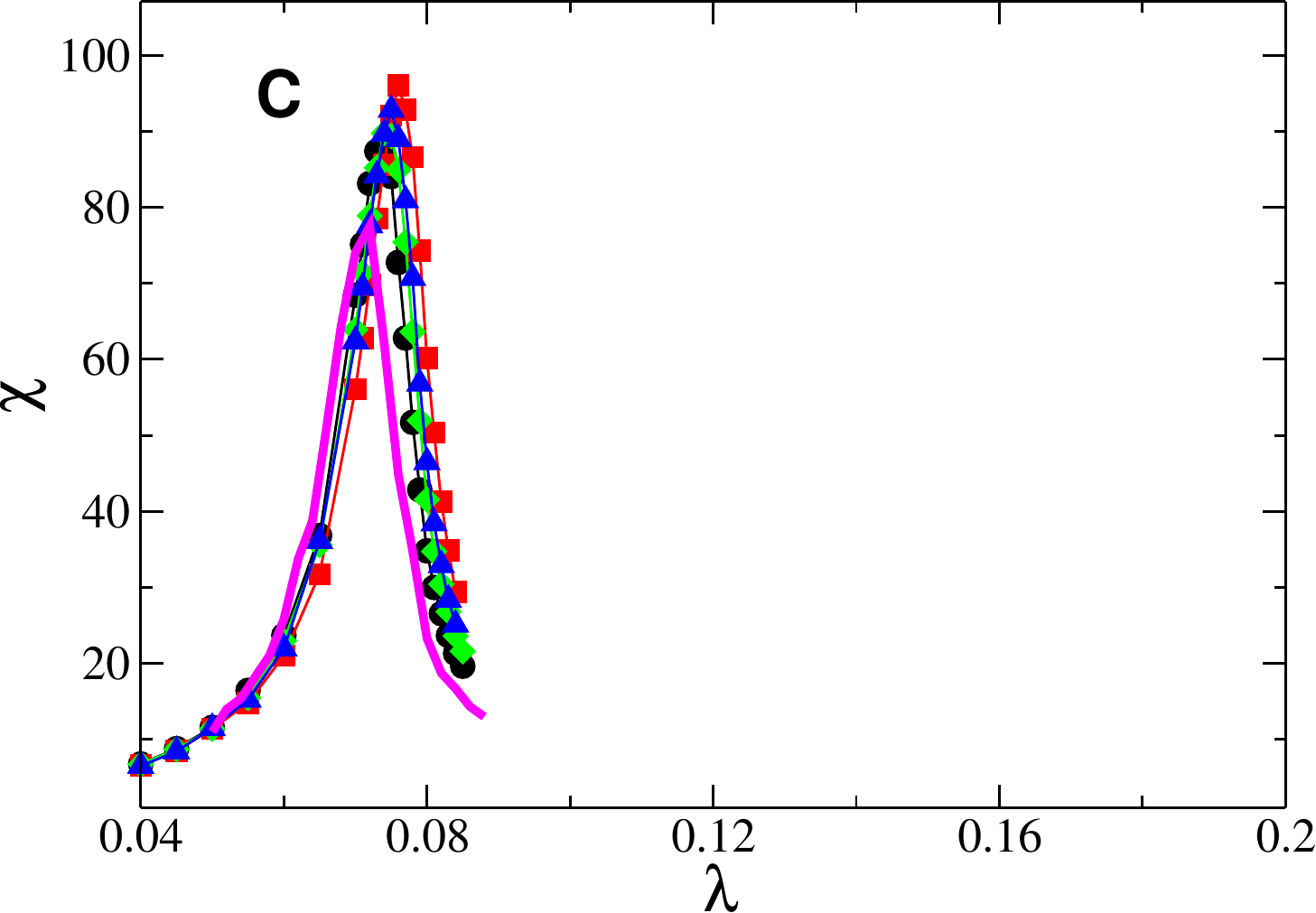} ~~ 
	\includegraphics[width=0.405\linewidth]{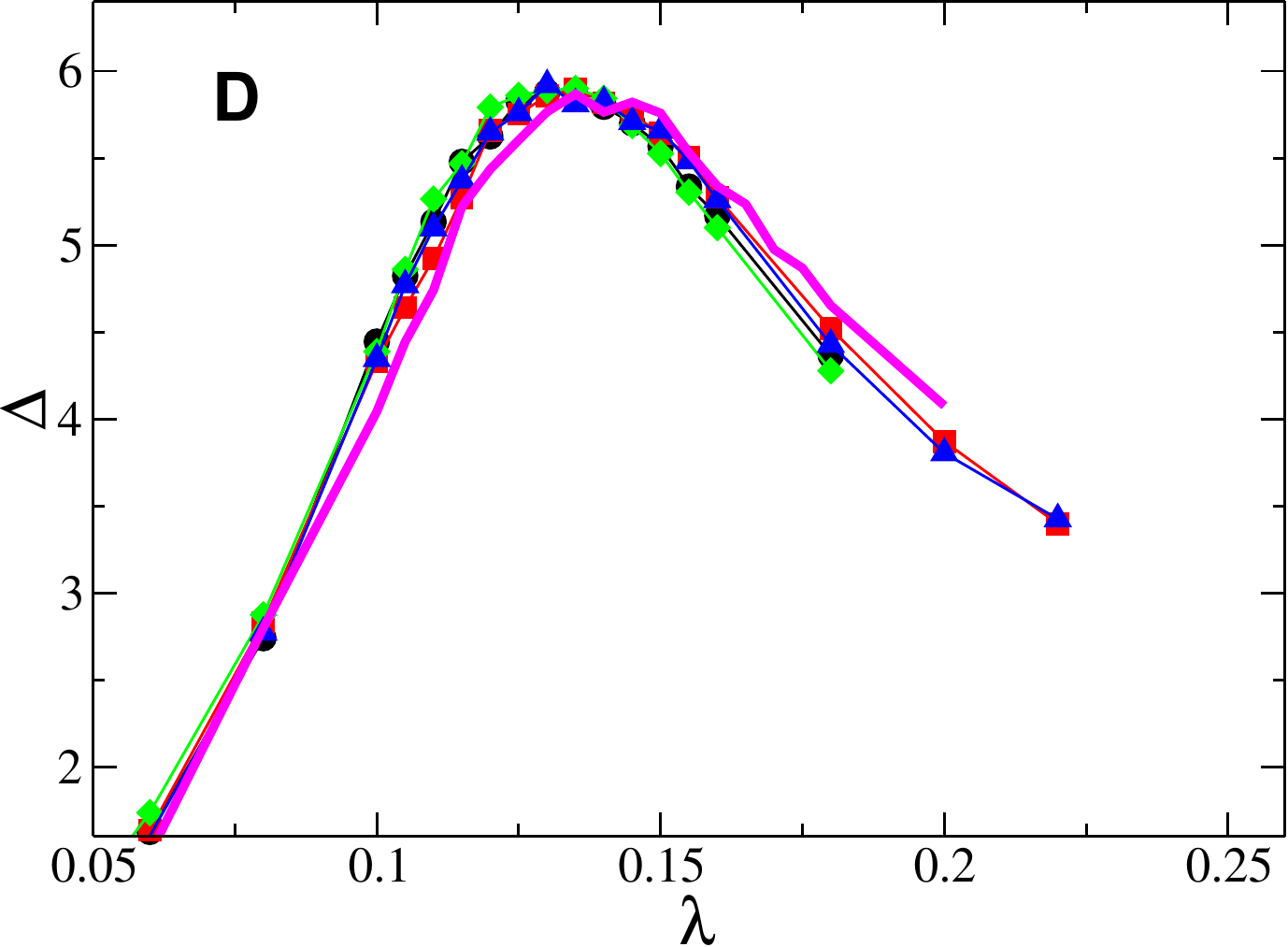}
	\caption{Effects of node and link intermittency on epidemic spreading over a scale-free backbone network. The susceptibility $\chi$ and variability $\Delta$ are shown as functions of the infection rate for (A,C) SIS and (B,D) SIR dynamics, respectively, with recovery rate $\mu=1$. Node intermittency (NI) is considered in panels (A,B), and link intermittency (LI) in panels (C,D). The backbone network has a power-law degree distribution with exponent $\gamma=2.75$, minimal degree $k_0=3$, and upper cutoff $k_\text{c}=2\sqrt{N}$ with $N=10^5$ for SIS and $N=10^4$ for SIR simulations. Thick magenta curves correspond to static networks in which $50\%$ of the nodes (A,B) or links (C,D) were randomly removed. The same horizontal-axis scale is used for the NI and LI simulations to facilitate direct comparison. }
	\label{fig:sis_sir_nodexlink}
\end{figure*}

The threshold is the central epidemic quantity that will be investigated. It can be estimated using the dynamical susceptibility~\cite{Sander2016,Cota2017} computed using the QS method in terms of the fluctuations of the epidemic prevalence after a relaxation period:
\begin{equation}
	\chi = N\frac{\langle{\rho^2}\rangle -\av{\rho}^2}{\av{\rho}}.
\end{equation}
The susceptibility exhibits a pronounced peak near the transition between disease extinction and sustained epidemic activity, providing an estimate of the epidemic threshold; see Figure~\ref{fig:sis_sir_nodexlink}. Since the QS method samples the long-time regime, the results are independent of the initial condition. We initialize the simulations with $50\%$ of randomly selected nodes infected.

Unlike SIS, the SIR model does not exhibit an active steady state: each outbreak eventually terminates when no infected individuals remain. We therefore perform independent outbreak simulations~\cite{Marro2005}, each starting from a single randomly selected infected individual. The relevant epidemic quantity is the attack rate $\rho_\textbf{R}$, defined as the final fraction of the population that has been infected and recovered. The epidemic threshold is estimated from the variability of the attack rate~\cite{Shu2015},
\begin{equation}
	\Delta = \frac{\sqrt{\av{\rho_\text{R}^2}-\av{\rho_\text{R}}^2}}{\av{\rho_\text{R}}},
\end{equation}
which also exhibits a peak near the epidemic threshold.

\section{Results}

We aim to isolate the effects of intermittency and of the corresponding waiting-time distributions. We therefore consider symmetric on/off dynamics, with $\Psi_\text{on}=\Psi_\text{off}=\Psi$ ($\beta_\text{on}=\beta_\text{off}=\beta$, and similarly for the other parameters). Consequently, on average, half of the nodes or links are active in the stationary regime. We set the recovery rate to $\mu=1$, which defines the time unit. For QS simulations of the SIS model, the dynamics is first relaxed for $t_\text{rlx}=5.10^5$, followed by an averaging period of $t_\text{av}=5.10^5$. For the SIR model, $10^5$ independent spreading simulations are initiated using a single infected node.

\subsection{Link versus node intermittency}

A simple reference scenario for the intermittent networks is obtained by randomly removing nodes or links from the backbone with probability $p_\text{off}=1-p_\text{on}$. For the symmetric case considered here, this corresponds to removing $50\%$ of these elements. Figure~\ref{fig:sis_sir_nodexlink} compares the SIS and SIR dynamics for different intermittency parameters with those on the corresponding diluted static networks. The static backbone network is generated with the uncorrelated configuration model~(UCM)~\cite{Catanzaro2005}, where an upper cutoff $k_\text{c}\sim \sqrt{N}$ guarantees the absence of degree correlations. For LI, the susceptibility and variability curves are essentially independent of the waiting-time distribution and peak very close to the corresponding peaks for the diluted static network. This shows that the effect of LI on epidemic spreading is largely explained by the average reduction in available contacts, with little additional influence from their temporal intermittency. 
This result, which is fully consistent with Ref.~\cite{Han2023a} when $\av{\tau}_\text{on}=\av{\tau}_\text{off}$ still with an additional non-Markovian infection process, is of limited interest by itself, since it essentially recovers the simple expectation based on the average reduction in contact availability. These results conform with the possibility of reducibility of non-Markovian dynamics to Markovian dynamics with effective epidemic rates~\cite{Han2023a,Han2023b,Starnini2017}.

For node intermittency, shown in Figure~\ref{fig:sis_sir_nodexlink}(a) and (b), the results are more complex and, consequently, far more interesting. Randomly removing $50\%$ of the nodes from the static backbone leads to approximately the same epidemic threshold as randomly removing $50\%$ of the links. This is expected because the nodes that remain in the network also lose, on average, $50\%$ of their contacts.\footnote{This argument can be generalized to arbitrary dilution. If nodes are retained with probability $p$, an edge remains functional only if both of its endpoints are retained, which occurs with probability $p^2$. The original backbone has $E=N\av{k}/2$ edges, while after dilution $E_\text{dil}=p^2E$ and $N_\text{dil}=pN$. Therefore, the average degree of the diluted network is $\av{k}_\text{dil}=2E_\text{dil}/N_\text{dil}=p\av{k}$. Thus, random removal of a fraction $1-p$ of the nodes produces the same average degree among the surviving nodes as random removal of the same fraction of links.}
Node intermittency, however, produces a strikingly different result. Both the susceptibility and the variability depend strongly on the inter-event-time distribution, and the SIS epidemic threshold increases by up to $60\%$ relative to the corresponding diluted static network, or by approximately $220\%$ relative to the undiluted static backbone. Thus, the suppression of epidemic spreading produced by node intermittency goes far beyond what can be explained by the average reduction in contact availability alone. A similarly strong effect is observed for the SIR model.

Figure~\ref{fig:sis_sir_nodexlink} shows that intermittently suppressing a fraction of the contacts is much less effective in hindering epidemic spreading than intermittently isolating individuals, the latter being closely related to quarantine strategies of direct epidemiological relevance. Interestingly, this difference cannot be explained by the total amount of contact activity experienced by each individual. To see this, we define the time-integrated contact activity of node $i$ over a sufficiently long time window $T$ as
\begin{equation}
	T_i = \sum_j \int_{t_0}^{t_0+T} A_{ij}(t) \, dt,
\end{equation}
where $A_{ij}(t)$ is the temporal adjacency matrix, with $A_{ij}(t)=1$ when the link between nodes $i$ and $j$ exists and is active, while $A_{ij}(t)=0$ otherwise. In the symmetric case considered here, LI and NI yield, on average, the same time-integrated contact activity: under LI, approximately half of the contacts of a node are available at any given time, whereas under NI all of its contacts are simultaneously available approximately half of the time.

The mechanism underlying this much stronger suppression produced by NI, therefore, involves how transmission opportunities are organized among individuals rather than simply their total availability. Under LI, an infected node typically retains a fraction of its active contacts and can therefore continue transmitting while other links are inactive. 
Under NI, by contrast, an inactive infected node is completely disconnected from its neighbors, simultaneously suppressing all of its transmission channels. Crucially, recovery proceeds independently of whether the node is active or inactive. Therefore, if an infected node remains inactive for a time comparable to or longer than the characteristic recovery time $\mu^{-1}$, it has a substantial probability of recovering before becoming active again. In fact, for an inactive period of duration $\tau_\text{off}$, this probability is $1-\exp(-\mu\tau_\text{off})$ since the recovery is Markovian. Thus, inactive periods can consume a substantial fraction, or even the totality, of the infectious period without providing any opportunity for transmission. This competition between recovery and the simultaneous suppression of all transmission channels provides a mechanism by which NI can inhibit epidemic spreading much more efficiently than LI, even when their time-integrated contact activities $T_i$ are the same.

\subsection{Intermittency distribution versus epidemic detriment}

\begin{figure*}[th]
	\centering
	\includegraphics[width=0.4\linewidth]{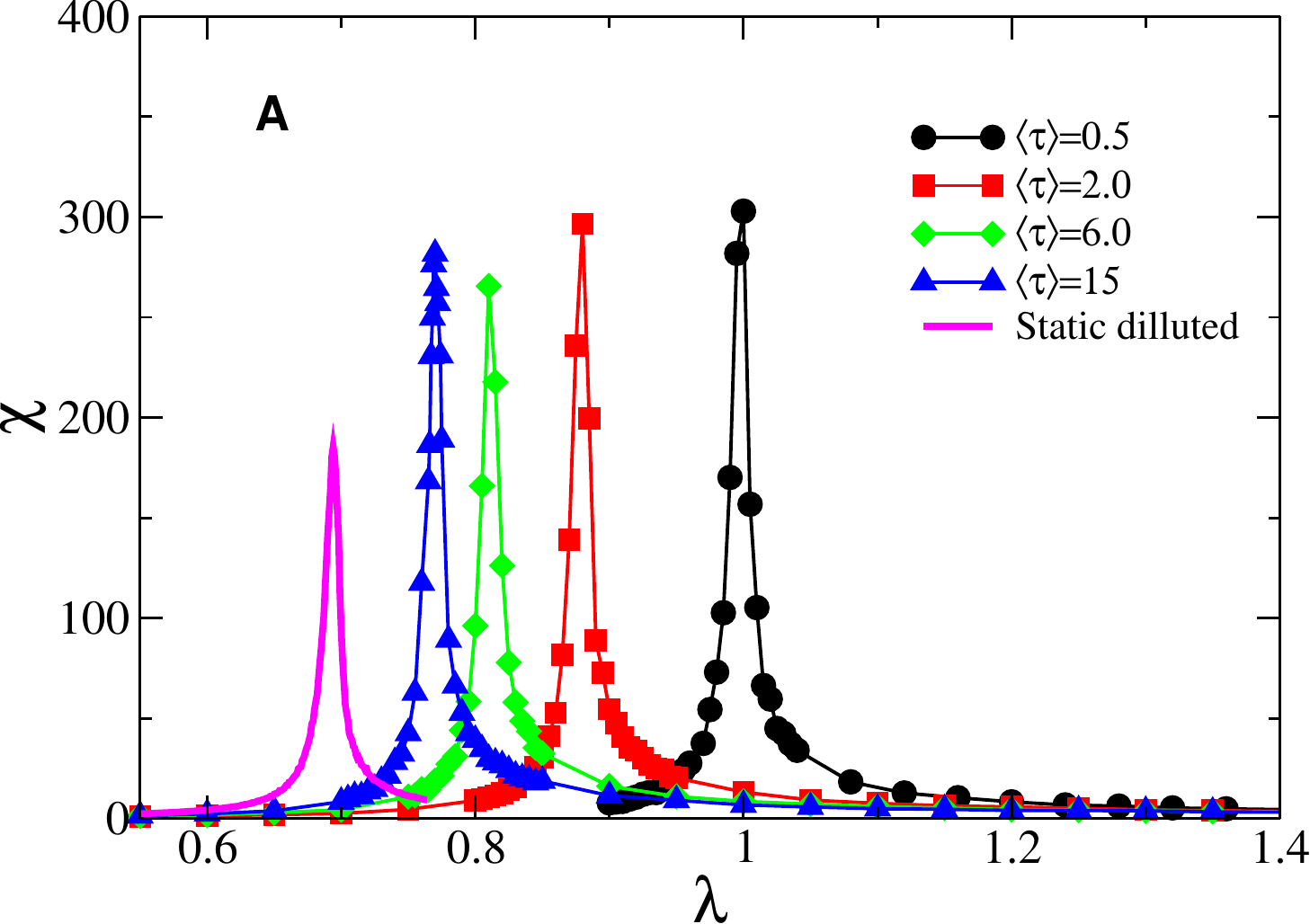}~
	\includegraphics[width=0.4\linewidth]{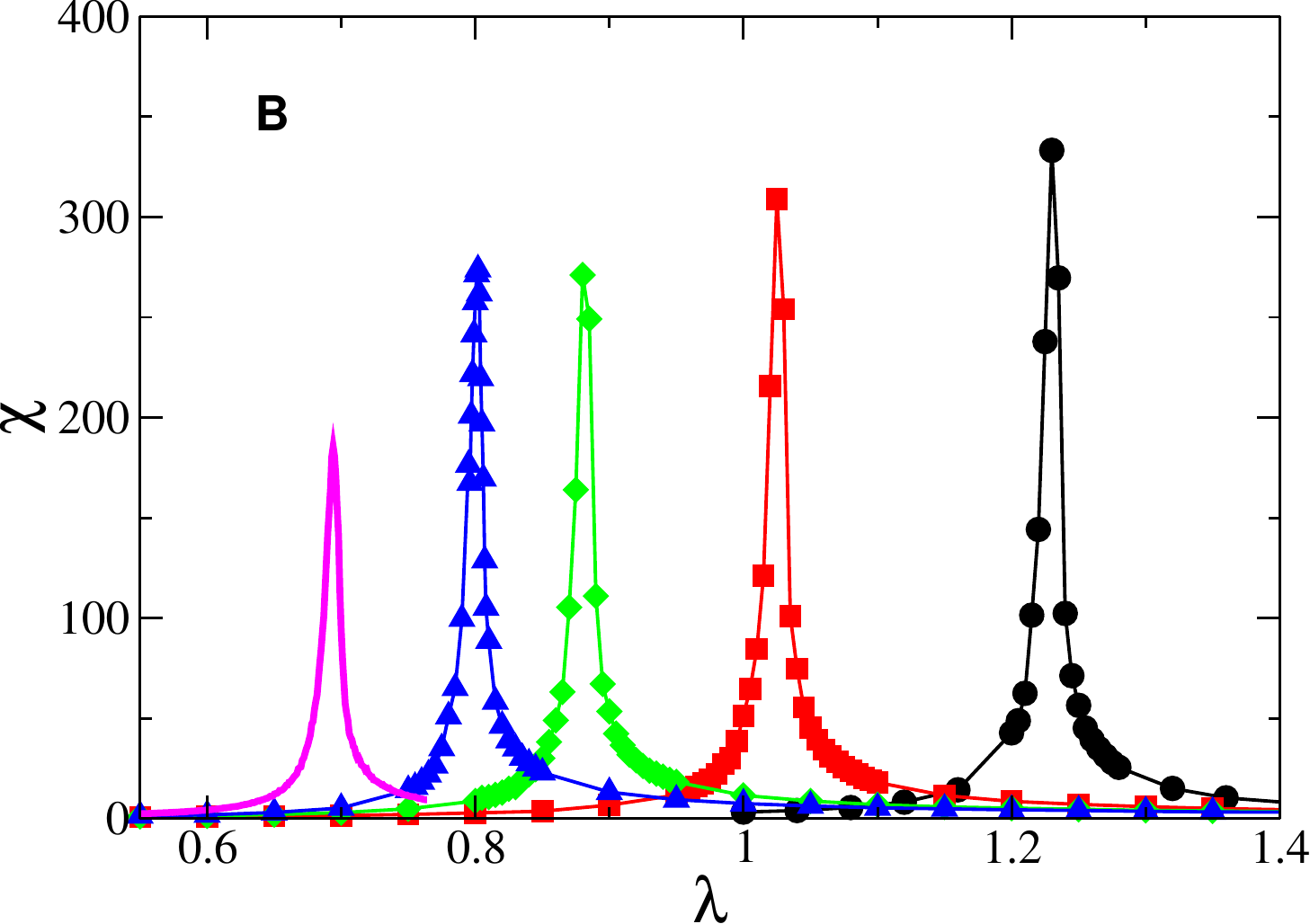}
	\caption{Intermittent networks with power-law inter-event-time distributions governing the switching between on and off states. Susceptibility curves are shown for (A) a heavy-tailed distribution with divergent variance, $\beta=2.3$, and (B) a distribution with finite variance, $\beta=3.8$. The backbone is a random regular network with degree $k=4$ and size $N=10^5$. The thick magenta curve corresponds to the static case in which $50\%$ of the nodes are permanently removed. }
	\label{fig:sisrrnnibetas}
\end{figure*}

Figure~\ref{fig:sis_sir_nodexlink} also shows that the properties of the inter-event-time distribution strongly affect epidemic suppression, as reflected in the displacement of the epidemic threshold. To disentangle these temporal effects from the structural heterogeneity of the backbone, we consider random regular networks (RRNs), in which all nodes have the same degree $k$ and connections are otherwise random~\cite{Bollobas2001}. For a fixed inter-event-time exponent $\beta$, epidemic spreading is progressively suppressed as the average switching time is reduced {in the range $\av{\tau}\ge 0.5$}, as systematically shown in Figures~\ref{fig:sisrrnnibetas}(a) and (b). The heterogeneity of the switching times also plays an important role. Increasing $\beta$, and therefore reducing the heterogeneity of the inter-event-time distribution, further increases the epidemic threshold, as can be seen by comparing the susceptibility peaks for $\beta=2.3$ and $\beta=3.8$ in Figure~\ref{fig:sisrrnnibetas}(a) and \ref{fig:sisrrnnibetas}(b) for a same average time $\av{\tau}$. In the opposite limit, as the average switching time becomes very large, the epidemic threshold approaches that of the diluted static network. This behavior is expected because nodes remain frozen in either the on or off state for increasingly long periods, making the network effectively static over the timescale of the epidemic dynamics. Similarly, heavier-tailed inter-event-time distributions, for which long switching intervals occur more frequently, shift the epidemic threshold toward the static-dilution limit. Overall, these results show that longer and more heterogeneous switching times facilitate epidemic spreading relative to switching node activity in smaller time scales. {The limit of $\av{\tau}\ll 1$ is of limited practical interest and is not discussed.}

The role of inter-event-time distributions in epidemic spreading has also been investigated by Masuda and Holme~\cite{Masuda2020} using SIS and SIR dynamics with renewal processes associated with the nodes of a backbone network. They reported that the epidemic outcome is not primarily controlled by the heavy tail or by the overall heterogeneity of the inter-event-time distribution, but rather by its probability mass at short inter-event times, with sufficiently frequent short intervals facilitating epidemic propagation. Their results therefore differ from the behavior observed here. This difference can be understood from the distinct roles played by the renewal processes in the two models. In Ref.~\cite{Masuda2020}, each renewal event activates a node, which then selects a neighbor and creates an instantaneous opportunity for transmission. Consequently, short inter-event times can generate rapid sequences of transmission opportunities. In our model, by contrast, renewal events determine the switching between finite on and off periods. Thus, the inter-event times in the two models control fundamentally different dynamical mechanisms, and their effects on epidemic spreading need not be the same. 

\subsection{Node intermittency versus  epidemic localization}

The SIS dynamics on heterogeneous networks can be strongly influenced by epidemic localization~\cite{Silva2020,Boguna2013}, whereby a relatively small subset of highly connected individuals (hubs) plays a disproportionate role in initiating and sustaining transmission throughout the network.\footnote{This does not imply that only a small fraction of the population is infected during the active epidemic phase. The epidemic can involve a finite fraction of the population while its persistence is mediated by a subset containing only a small fraction of the individuals.} The mechanism by which these hubs sustain the epidemic depends on how they are organized within the network. If hubs form a densely connected core, they can repeatedly infect one another, maintaining a persistent source of infection that subsequently spreads to the rest of the population. If, instead, hubs are sparsely distributed throughout the network, direct transmission between them is unlikely. In this case, infection can persist locally through repeated transmission between a hub and its many neighbors: the infected hub transmits the disease to several neighbors, some of which may remain infected long enough to reinfect the hub after it recovers~\cite{Boguna2013}. This local feedback can maintain infection around a hub for a sufficiently long time to allow occasional transmission through paths of less-connected individuals to reach another hub. Although such long-range transmission events are relatively rare, they provide a mechanism for infection to propagate between otherwise weakly connected regions of sustained epidemic activity.

From a theoretical perspective, understanding how intermittency affects epidemic localization is crucial for explaining the strong suppression produced by NI. We investigate SIS dynamics on UCM networks of different sizes. For $\gamma=2.25$, where epidemic activity is known to be sustained by a densely connected core of highly connected nodes~\cite{Castellano2012}, the epidemic thresholds for both LI and NI exhibit the same dependence on network size and asymptotically approach zero. This indicates that the collective activation mechanism provided by the densely connected core is sufficiently robust to persist in the presence of intermittency.

\begin{figure*}[t!]
	\centering
	\includegraphics[width=0.4\linewidth]{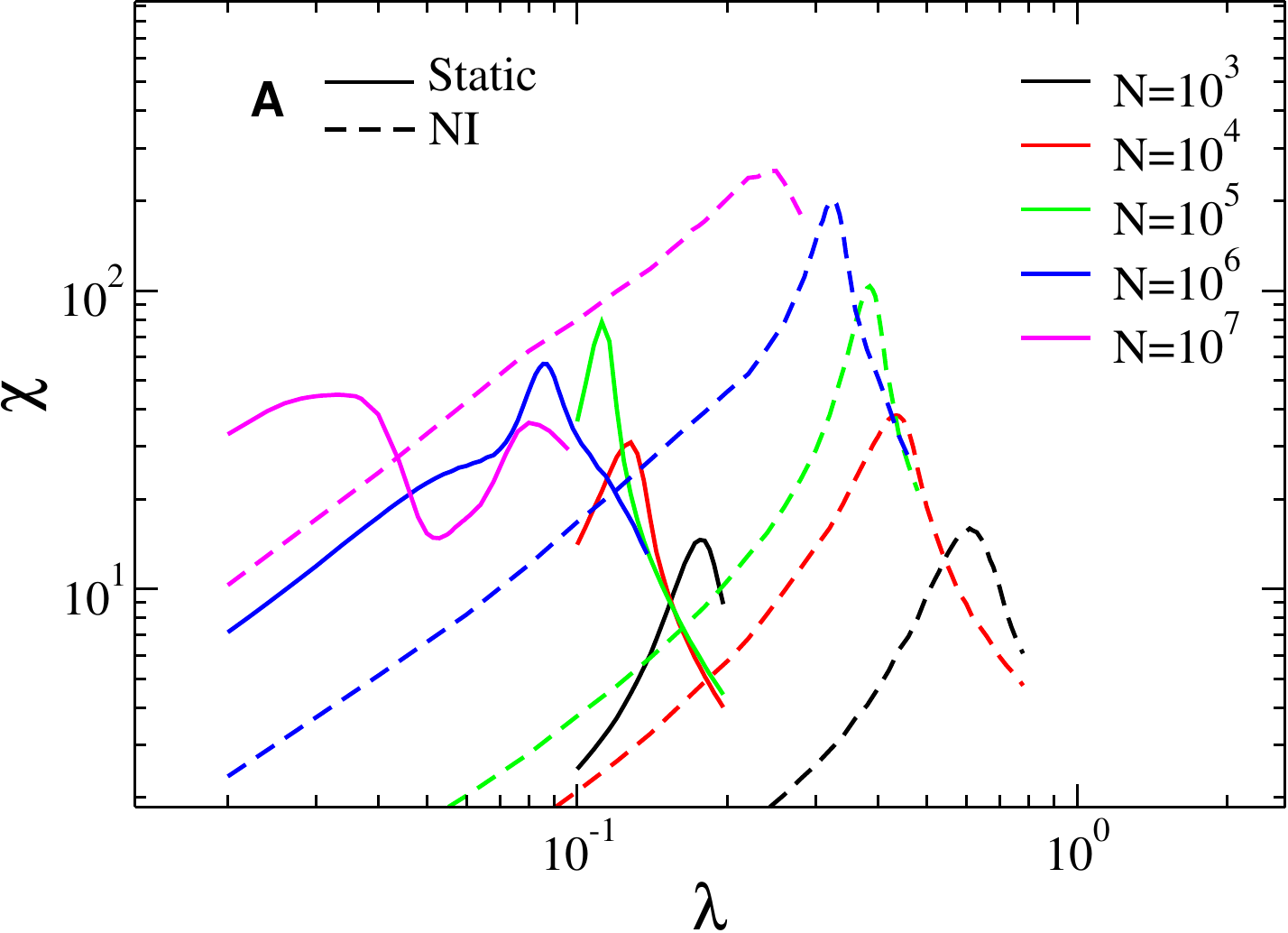}~~
	\includegraphics[width=0.4\linewidth]{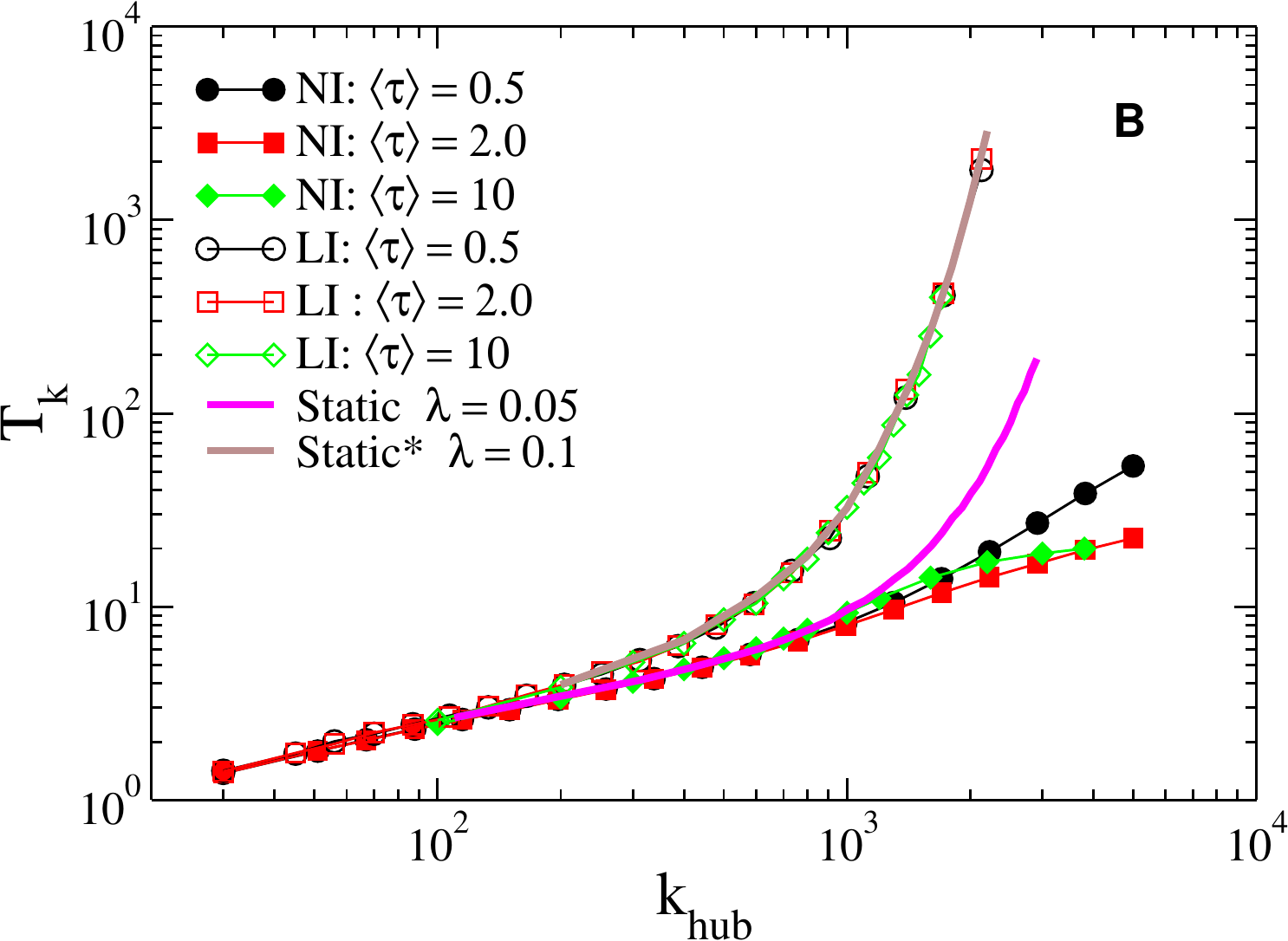}
	
	\caption{Localization effects for epidemic on intermittent networks.The inter-event-time has exponent $\beta=3.7$ and average inter-event time $\av{\tau}=0.5$. (A) Susceptibility curves for SIS dynamics on power-law networks with degree exponent $\gamma=3.5$, considering static networks (solid lines) and node intermittency (dashed lines). (B) Epidemic lifespan for SIS dynamics on star graphs consisting of a central node connected to $k$ leaves, considering node (solid lines) and link (dashed lines) intermittency, with infection rate $\lambda=0.1$. The thick magenta solid line corresponds to simulations on static star graphs with the infection rate reduced by a factor of two, while the thick brown line corresponds to static star graphs with half the size.
    \vspace{-0.2cm}}
	\label{fig:sisnigamma3p5}
\end{figure*}

More interesting effects emerge for $\gamma>3$, where the epidemic persistence relies on localized reinfection around individual hubs. Figure~\ref{fig:sisnigamma3p5}(a) compares the susceptibilities obtained for the static network and under NI. For sufficiently large static networks, the susceptibility exhibits multiple peaks, reflecting the successive activation of different localized regions and providing a characteristic signature of epidemic localization around the most connected nodes~\cite{Mata2015}. Under NI, however, the localized activation associated with the most connected hub is no longer observed. Node intermittency therefore profoundly alters the localization pattern underlying hub-driven epidemic spreading, breaking the local feedback between a hub and its neighbors that would otherwise sustain infection for long periods.

The effect of intermittency on hub localization can be better understood by considering epidemic dynamics on star graphs, consisting of a central hub connected to $k$ neighbors of degree one. This simple structure isolates the local hub--neighbor feedback mechanism that sustains infection around sparsely distributed hubs in heterogeneous networks~\cite{Boguna2013}. We compute the epidemic lifespan starting with the hub infected and all its neighbors susceptible. For a static star graph, the epidemic lifespan increases exponentially with the hub degree as $T_k\sim e^{a\lambda^2 k}$~\cite{Boguna2013}, where $a$ is a constant. Thus, sufficiently large hubs can maintain infection locally for very long times.

Figure~\ref{fig:sisnigamma3p5}(b) shows the epidemic lifespan for SIS dynamics on star graphs with $k$ leaves under LI and NI, considering symmetric on and off inter-event-time distributions. Under LI, the epidemic lifespan matches that obtained for a static star graph with half as many links. This confirms the picture discussed above: the main effect of LI can be accounted for by the average fraction of inactive links. For NI, we instead compare with a static star graph of the same size but with the infection rate reduced by a factor of two, accounting for the fact that the hub is active, on average, only $50\%$ of the time. The NI results initially follow this effective static approximation remarkably well, up to approximately $k=10^3$, where the static case begins to display its characteristic exponential increase. Beyond this regime, however, the NI lifespan continues to increase much more slowly. Node intermittency therefore changes not only the magnitude of the epidemic lifespan but also strongly suppresses highly connected individuals from acting as long-lived reservoirs of infection.

This qualitative change in scaling can have important consequences in the limit of very large networks. A related result was obtained for the SIRS model, in which recovered individuals become susceptible again at rate $\alpha$: the epidemic lifespan around a hub was found to increase algebraically rather than exponentially with its degree~\cite{Ferreira2016a} (see also~\cite{Friedrich2024} for a rigorous treatment). This result led to the conjecture that isolated hubs cannot sustain endemic activity at arbitrarily small infection rates in the infinite-network limit, in contrast with the well-established behavior of SIS dynamics on sufficiently heterogeneous networks~\cite{Pastor-Satorras2015,Boguna2013} (see~\cite{Chatterjee2009,Mountford2013} for rigorous treatments). The slow growth observed here suggests that NI may produce an analogous qualitative change by disrupting the hub-localized mechanism responsible for sustaining SIS dynamics. Our simulations, however, cannot establish the asymptotic behavior for arbitrarily large star graphs or networks. We therefore leave as a conjecture that NI may alter the infinite-size limit of SIS dynamics by preventing isolated hubs from sustaining epidemic activity at arbitrarily small infection rates.

\section{Conclusion}
The dynamics of infectious diseases is shaped not only by the heterogeneous structure of contact networks but also by the temporal organization of interactions. Human activity is intermittent and often characterized by broadly distributed inter-event times~\cite{Karsai2018}, motivating temporal-network descriptions of epidemic spreading~\cite{Masuda2020}. Here, we investigated this problem on a fixed backbone network in which either links or nodes alternate between active and inactive states according to renewal processes, while the epidemic dynamics itself remains Markovian.

Our main result is that link intermittency (LI) and node intermittency (NI) affect epidemic spreading in fundamentally different ways. LI can be effectively described by a suitable rescaling of the transmission rate, indicating that its main effect is the reduction in available contacts. NI, in contrast, raises the epidemic threshold substantially beyond what is expected from the same average reduction in contact activity. The difference arises because, under NI, all transmission pathways of an inactive individual are simultaneously interrupted while recovery continues. An infected individual can therefore recover during an inactive period without transmitting the pathogen.

Node intermittency also strongly affects epidemic localization. While the collective activation of densely connected cores remains robust against intermittency, NI disrupts the infection--reinfection cycles that sustain epidemic activity around isolated hubs. On star graphs, this effect changes the growth of the epidemic lifespan with hub degree from the exponential behavior of static SIS dynamics to an approximately algebraic dependence within the range investigated. This suggests that NI may alter the asymptotic hub-activation mechanism of SIS dynamics, although this remains a conjecture requiring further theoretical investigation.

The temporal organization of NI is also relevant:{ for the investigated distributions of inter-event times}, shorter and less heterogeneous inter-event times lead to higher epidemic thresholds and stronger suppression of spreading. These findings have implications for quarantine and temporary-isolation strategies. For the same average reduction in contact activity, concentrating the interruption on individuals can be substantially more effective than distributing it among individual contacts. Moreover, our results suggest that frequent and more regularly distributed periods of isolation can be more effective than longer and highly heterogeneous ones. Although our model does not reproduce the complexity of real quarantine policies, it highlights a general principle: epidemic control depends not only on how much contact activity is reduced, but also on how this reduction is distributed among individuals and organized in time.

\appendix

\section{The algorithm}
\label{app:algorithm}

\newcommand{\Ninf}{N_\text{inf}}
\newcommand{\NIS}{N_\text{IS}}

The network state is initialized by simulating the activation--deactivation dynamics for a sufficiently long time to reach the stationary regime. For LI, each link is labeled as on or off, and the time $T_{ij}$ of its next switching event is stored. Similarly, for NI, each node is labeled as on or off, together with the time $T_i$ of its next switching event.

Epidemic spreading is simulated using an adaptation of the optimized Gillespie algorithm for static networks~\cite{Cota2017}. A list $\mathcal{I}$ of infected nodes is maintained and continuously updated, together with the number of infected individuals $\Ninf$ and the total number of stubs emanating from them, $\NIS=\sum_{i\in\mathcal{I}}k_i$. At each step, a recovery event is selected with probability
	$p={\mu\Ninf}/{(\mu\Ninf+\lambda \NIS)}$,
in which an infected node $i\in\mathcal{I}$ is chosen uniformly at random and becomes susceptible in the SIS model or recovered in the SIR model.

With complementary probability $1-p$, an infection attempt is performed. An infected node $i\in\mathcal{I}$ is selected with probability proportional to its degree $k_i$, and one of its neighbors $j$ is then chosen uniformly at random. For LI, the state of the link $(i,j)$ is updated whenever necessary by running its renewal process until $t<T_{ij}$. If the link is active and $j$ is susceptible, $j$ becomes infected; otherwise, the attempt produces no change in the epidemic state.

For NI, the activity state of the selected infected node $i$ is first updated by running its renewal process until $t<T_i$. If $i$ is active, a neighbor $j$ is selected uniformly at random and its activity state is similarly updated until $t<T_j$. Infection occurs only if both $i$ and $j$ are active and $j$ is susceptible. Otherwise, the attempt produces no change in the epidemic state.

After every event or unsuccessful infection attempt, the simulation time is incremented by
	$\delta t= {-\ln u}/{(\mu\Ninf+\lambda \NIS)}$,
where $u$ is a random number uniformly distributed in $(0,1)$.

\begin{acknowledgments}{S.C.F  acknowledge the financial support by the  \textit{Fundação de Amparo à Pesquisa do Estado de Minas Gerais} (FAPEMIG)-Brazil (Grant No. APQ-01973-24) \textit{Conselho Nacional de Desenvolvimento Científico e Tecnológico} (CNPq)-Brazil (Grants No. 407871/2025-0 and 310984/2023-8),  and \textit{Fundação de Amparo à Pesquisa do Estado de São Paulo} FAPESP (Grant No. 25/24366-1). J. T. M. thanks FAPESP (Grant No. 2026/09600-0). This study was financed in part by the Coordenação de Aperfeiçoamento de Pessoal de Nível Superior - Brasil (CAPES) - Finance Code 001.}
\end{acknowledgments}

\bibliography{intermittent}

\end{document}